\documentclass[aps,pra,preprint,groupedaddress,showpacs]{revtex4}
\usepackage{graphicx}
\usepackage{bm}
\begin{document}
\preprint{Version 1.1}

\title{Calculation of Radiative Corrections to E1 matrix elements \\
in the Neutral Alkalis}

\author{J. Sapirstein}
\email[]{jsapirst@nd.edu}
\affiliation{
Department of Physics, University of Notre Dame, Notre Dame, IN 46556}

\author{K. T. Cheng}
\email[]{ktcheng@llnl.gov}
\affiliation{
University of California, Lawrence Livermore National Laboratory,
Livermore, CA 94550}

\begin{abstract}
Radiative corrections to E1 matrix elements for $ns-np$ transitions
in the alkali metal atoms lithium through francium are evaluated.
They are found to be small for the lighter alkalis but significantly
larger for the heavier alkalis, and in the case of cesium much larger
than the experimental accuracy. The relation of the matrix element
calculation to a recent decay rate calculation for hydrogenic ions
is discussed, and application of the method to parity nonconservation
in cesium is described.
\end{abstract}


\maketitle

\section{Introduction}
High accuracy measurements of atomic lifetimes are difficult, with only
a few examples of precisions well under one percent known. For this
reason, consideration of radiative corrections to lifetimes is usually
not necessary. Until recently the most prominent exception to this
situation was the decay rates of orthopositronium and parapositronium,
where determinations of accuracy 180 ppm \cite{Vallery} and 215 ppm
\cite{Gidley} respectively have been made. In these exotic atoms the
radiative corrections start in order $\alpha$ with large coefficients,
and the leading radiative correction are clearly visible, both because
in these two-body atoms the wave function is known analytically and
because quantum electrodynamic (QED) corrections start in order $\alpha$
with large coefficients, giving contributions of 2.1 percent and 0.5 percent 
respectively.

In recent years, however, a new approach has been developed that exploits
the fact that the dipole-dipole potential between two alkali atoms,
which goes as as $C_3/R^3$, can be accurately measured, and $C_3$ is
proportional to the lifetimes of $np$ states of the atoms. This has
allowed the determination of the lifetime of the $2p_{1/2}$ state of
lithium as 27.102(7) ns \cite{liexp}, the lifetime of the $3p_{3/2}$
state of sodium as 16.230(16) ns \cite{naexp}, the lifetimes of the
$4p_{1/2}$ and $4p_{3/2}$ states of potassium as 26.69(5)ns and 26.34(5) ns
\cite{kexp}, the lifetimes of the $5p_{1/2}$ and $5p_{3/2}$ states of
rubidium as 27.75(8) ns and 26.25(8) ns \cite{rbexp}, and the lifetimes
of the $6p_{1/2}$ and $6p_{3/2}$ states of cesium as 34.88(2) ns and
30.462(3) ns \cite{csexp}. (References to other experimental determinations
of lifetimes, some of which are of higher accuracy, can be found in the
above references. We note also that the lifetimes of the francium $7p_{1/2}$
and $7p_{3/2}$ states have been measured as 29.45(11) ns and 21.02(11) ns
using a different technique \cite{frexp}.) These high accuracies, which
for matrix elements correspond at the best to 50 ppm, now make the
calculation of radiative corrections of interest, even though unlike the
case of positronium the corrections to E1 matrix elements are known to
enter, for hydrogenic ions \cite{SPC}, in order $\alpha (Z\alpha)^2$.

There are two other reasons for carrying out calculations of these
radiative corrections. Firstly, this is a relatively unexplored region
of QED. It has only been very recently that the first full calculations
of one-loop radiative corrections to the decay rate of the hydrogen
isoelectronic sequence have been carried out \cite{SPC}. That calculation
was done by considering the imaginary part of the two-loop Lamb shift,
which is equivalent to calculating the shift in the lifetime. Here we
adopt a different method, and instead calculate radiative corrections
to the associated transition matrix element. The introduction of this
technique requires a nontrivial modification of the previous formalism,
and also provides a check of that method.

The second additional reason for calculating radiative corrections
to transition matrix elements in the alkalis is the interest in parity
nonconservation (PNC) in cesium \cite{Wieman}. Recently a large binding
correction to the Z boson-electron vertex radiative correction has
been found that has significant implications for the standard model.
The lowest-order radiative correction, $-\alpha/2\pi$, has been
shown \cite{pncrad1,pncrad2,pncrad3} to be enhanced through binding
corrections by almost an order of magnitude. However, the actual
radiative correction is to the E1 matrix element of a $6s$ electron
to a $7s$ electron, with an opposite parity component present in one
of these electrons induced by Z exchange with the nucleus. The full
radiative correction calculation needed then involves the evaluation
of the diagrams in Fig.~\ref{fig:fig1}, of which diagram~\ref{fig:fig1}c,
where the radiative correction is on the photon rather than the Z vertex,
is of the kind to be treated here. Thus the calculations on cesium that
will be presented here will be of use for this larger scale task.

We note that the methods used in this paper have previously been applied
to calculating the Lamb shift \cite{neutrallamb} and the radiative
correction to hyperfine splitting \cite{HFS} for the ground states of
the alkalis, where QED effects larger than experimental accuracy were
found. However, the calculations remain untested because of the relative
inaccuracy of many-body methods (with the exception of lithium, where
highly accurate variational methods are available \cite{liv}). We will
see that the same situation is present for alkali lifetimes, and thus
the present work provides yet another impetus to many-body theory to
reach the accuracies presently of interest for both radiative
corrections and experiment.

In the next section, the generalization of the method used for the
previous calculations of radiative corrections in the alkalis will
be laid out. Of particular interest are certain issues related to the
fact that the Gell-Mann-Low formalism for energies used in the past
work has to be changed because we are now instead dealing with matrix elements.
In Section \ref{sec:hions}, the technique is applied to hydrogenic ions
to compare with previous work, and in Section \ref{sec:results} the main
results of the paper are presented. We conclude with a discussion of how
this approach can be made more accurate and how it can be generalized for
application to the calculation of radiative corrections to cesium PNC.

\section{Formalism}
When one is interested in calculating energy shifts in atoms, Sucher's
generalization \cite{Sucher} of the Gell-Mann-Low formalism provides
a systematic way to derive them from the S-matrix through the formula
\begin{equation}
\Delta E = \lim_{\epsilon \to 0, \, \lambda \to 1}
    { i \epsilon \over 2 } \,
    { 1 \over S_{\epsilon,\lambda} } \,
    {\partial S_{\epsilon,\lambda} \over \partial \lambda},
\label{eq:eq1}
\end{equation}
where $S_{\epsilon, \lambda}$ is the S-matrix with the interaction
Hamiltonian $H_I(t)$ multiplied by the factor
\begin{equation}
\lambda e^{-\epsilon |t|}.
\end{equation}
The overall factor of $\epsilon$ is compensated by the fact that the
S-matrix diverges as ${1 / \epsilon}$. In higher orders, factors
of ${1 / \epsilon^2}$ are encountered in the numerator that are
canceled by $1/\epsilon$ terms coming from expanding the denominator.
When considering matrix elements the factor $\lambda$, which accounts
for a combinatorial factor specific to the Gell-Mann-Low formalism,
will not be used, but we continue to use the exponential damping
factor $\epsilon$, which will lead to the frequent occurrence of
what is effectively the delta function,
\begin{equation}
D_{\epsilon}(x) = {1 \over \pi} \, {\epsilon \over x^2 + \epsilon^2}.
\end{equation}

We replace the time-independent perturbation of Ref.~\cite{BCS} with
a time dependent Hamiltonian in the length-gauge form appropriate for
describing the absorption of a photon in a electromagnetic field of
strength $E_0$ linearly polarized in the $z$ direction,
\begin{equation}
H_{\rm LG} = e E_0 \!\int\! d^{\,3}x \, \psi^{\dagger}(\vec x, t)
    \,\vec x \cdot \!\hat{z} e^{-i \omega t} \,
    \psi(\vec x,t) a(\vec k, \hat{z}).
\end{equation}
We have made the dipole approximation, a good approximation for the
neutral alkalis, so the momentum of the initial photon, $\vec k$,
plays no role in the following. In addition we suppress the factor
$e E_0$ in the following. We consider the matrix element of this
Hamiltonian between an initial state $v$ taken to be a $ns$ ate
($2s$ for lithium, $3s$ for sodium, etc.) along with a photon with
energy $\omega$, and a final state $w$ taken to be a $np_{1/2}$ or
$np_{3/2}$ state with the same principal quantum number $n$. It is
important to keep the photon energy $\omega$ distinct from the
resonance energy $\omega_0 \equiv \epsilon_w - \epsilon_v$, and
in particular the limit $\epsilon \to 0$ is always understood to
be taken before the limit $\omega \to \omega_0$. In lowest order,
the S-matrix is then
\begin{equation}
S = -2 \pi i  D_{\epsilon}(\omega - \omega_0) z_{wv},
\end{equation}
and following the convention just mentioned, this becomes the usual
\begin{equation}
S = [ - 2 \pi i \delta(\omega - \omega_0) ] z_{wv}.
\end{equation}
In the following, we shall suppress the factor in square brackets.
In Table \ref{tab:tab1}, we present results for the dipole matrix
element $r_{wv}$, with
\begin{equation}
r_{wv} = {\int}_0^{\infty} dr\;r\;\Big[g_w(r) g_v(r) + f_w(r) f_v(r)\Big],
\end{equation}
from which $z_{wv}$ can be obtained by either multiplying by a factor of
$(1/3) (-1)^{j_w -m_v}$ or $\sqrt{2}/3$ for the $ns_{1/2} - np_{1/2}$ and
$ns_{1/2} - np_{3/2}$ transitions, respectively. These are lowest-order
results obtained using the local Kohn-Sham potential, a description
of which can be found in Ref.~\cite{neutrallamb}. Issues involved in
correcting these results with many-body methods will be addressed in
the concluding section, but here we will concentrate on radiative
corrections. In evaluating these we will work in terms of the ratio
$R_{wv}$ defined through
\begin{equation}
\delta z_{wv} \equiv {\alpha \over \pi} z_{wv} R_{wv},
\label{eq:rwv}
\end{equation}
such that
\begin{equation}
z_{wv} + \delta z_{wv} = z_{wv} \left(1 + {\alpha \over \pi} R_{wv}\right).
\end{equation}
At this point we also define the frequently occurring self-energy operator
\begin{equation}
\Sigma_{ij}(\epsilon) = - 4 \pi i \alpha
    \!\int\! d^{\,3}x
    \!\int\! d^{\,3}y
    \!\int\! {d^{\,n}k \over (2\pi)^n} \,
    {1 \over k^2 + i \delta} \,
    \bar{\psi}_i(\vec x) \gamma_{\mu}
    S_F(\vec x , \vec y; \epsilon - k_0) \gamma^{\mu} \psi_j(\vec y),
\end{equation}
in terms of which the lowest-order self-energy part of the Lamb shift
of a state $v$, treated in Ref.~\cite{neutrallamb}, is simply
$\Sigma_{vv}(\epsilon_v)$.

Most of the discussion of radiative corrections to $z_{wv}$ is very similar
to the treatment using the Gell-Mann-Low formalism given in Ref.~\cite{BCS},
except here we pull out a factor of $-2 \pi i \delta(\omega - \omega_0)$
as opposed to isolating $1/\epsilon$ terms. As described in more detail
in that paper, three diagrams shown in Fig.~\ref{fig:fig2} contribute to
the radiative correction to the matrix element. The S-matrix associated with
the vertex (V) diagram of Fig.~\ref{fig:fig2}b is given by
\begin{eqnarray}
S_{\rm V} &=& -32 \pi^4 \alpha
    \!\int\! d^{\,3}x
    \!\int\! d^{\,3}y
    \!\int\! d^{\,3}z
    \!\int\! {d^{\,n}k \over (2 \pi)^n} \,
    {e^{i \vec k \cdot (\vec x - \vec z)} \over k^2 + i \delta}
      \int\! {dE_1 \over 2 \pi}
    \!\int\! {dE_2 \over 2 \pi} \,
    \bar{\psi}_w(\vec x) \gamma_{\mu}
    S_F(\vec x, \vec y; E_1)
    \,\vec y \cdot \!\hat{z}\, \gamma_0 \,
\nonumber \\ && \times \,
    S_F(\vec y, \vec z; E_2) \gamma^{\mu} \psi_v(\vec z)
    D_{\epsilon}(E_1 + k_0 - \epsilon_w)
    D_{\epsilon}(E_2 + k_0 - \epsilon_v)
    D_{\epsilon}(E_2 - E_1 + \omega).
\end{eqnarray}
One can make the substitutions $E_1 \to \epsilon_w - k_0$ and
$E_2 \to \epsilon_v - k_0$ in the electron propagators, which
then allows the $E_1$ and $E_2$ integrations to be carried out,
\begin{eqnarray}
&&   \int\! {dE_1 \over 2 \pi}
    \!\int\! {dE_2 \over 2 \pi} \,
    D_{\epsilon}(E_1 + k_0 - \epsilon_w)
    D_{\epsilon}(E_2 + k_0 - \epsilon_v)
    D_{\epsilon}(E_2 - E_1 - \omega)
\nonumber \\ & = &
    {3 \epsilon \over 4 \pi^3} \,
    {1 \over (\omega - \omega_0)^2 + 9 \epsilon^2}
\nonumber \\ & \to &
    {\delta(\omega - \omega_0) \over 4 \pi^2}.
\end{eqnarray}
We note that in energy calculations where the factor $\omega - \omega_0$
vanishes, a factor 1/3 results that is canceled because a derivative with
respect to the factor $\lambda$ present in the energy formula acts on a
factor $\lambda^3$. Here the factor 1/3 is not present for a different
reason, that being the fact that the effective infinitesimal factor used to
obtain the energy conserving delta function is $3 \epsilon$. We then find,
after pulling out the factors mentioned above, a vertex contribution of
\begin{eqnarray}
\delta z_{wv}({\rm V}) &=& -4 \pi i \alpha
    \!\int\! d^{\,3}x
    \!\int\! d^{\,3}y
    \!\int\! d^{\,3}z
    \!\int\! {d^{\,n}k \over (2 \pi)^n} \,
    {e^{i \vec k \cdot (\vec x - \vec z)} \over k^2 + i \delta} \,
    \bar{\psi}_w(\vec x) \gamma_{\mu}
    S_F(\vec x, \vec y; \epsilon_w - k_0)
    \,\vec y \cdot \!\hat{z}\, \gamma_0 \,
\nonumber \\ && \times \,
    S_F(\vec y, \vec z; \epsilon_v -k_0) \gamma^{\mu} \psi_v(\vec z).
\end{eqnarray}
This expression has both ultraviolet divergences and reference state
singularities that cancel with the side graphs of Figs.~\ref{fig:fig2}a
and \ref{fig:fig2}c discussed below. The ultraviolet divergence is
isolated analytically by replacing both bound state propagators with
free propagators, which gives an ultraviolet divergent term along with
a finite remainder we tabulate as $R_{wv}({\rm V};00)$ in the first
row of Table~\ref{tab:tab2}. (We note at this point that we present
results only for $ns - np_{1/2}$ transitions. Those for $ns - np_{3/2}$
transitions should not be too different for neutral systems considered
here, as radiative corrections to the lifetimes of the $2p_{1/2}$ and
$2p_{3/2}$ states are essentially the same for low-$Z$ hydrogenic
ions \cite{SPC}.) We then form the ultraviolet finite difference
of $\delta z_{wv}({\rm V})$ and $\delta z_{wv}({\rm V};00)$ and
evaluate it in coordinate space. A Wick rotation $k_0 \to i \omega$
is carried out, which passes poles: the separate contributions are
tabulated as $R_{wv}({\rm V};i\omega)$ and $R_{wv}({\rm V;Poles})$
in the second and third rows of Table~\ref{tab:tab2}. The reference
state singularity mentioned above is present in the first term,
and is regulated through the replacement
$\epsilon_v \to \epsilon_v (1 - i \delta)$ and
$\epsilon_w \to \epsilon_w (1 - i \delta)$ with
$\delta$ typically chosen to be $10^{-6}$.

The treatment of the side graphs also differs from the previous energy
approach. The starting expression for the ``side-right'' (SR) diagram of
Fig.~\ref{fig:fig2}a is
\begin{eqnarray}
S_{\rm SR} &=& -32 \pi^4 \alpha
    \!\int\! d^{\,3}x
    \!\int\! d^{\,3}y
    \!\int\! d^{\,3}z
    \!\int\! {d^{\,n}k \over (2 \pi)^n} \,
    {e^{i \vec k \cdot (\vec y - \vec z)} \over k^2 + i \delta}
      \int\! {dE_1 \over 2 \pi}
    \!\int\! {dE_2 \over 2 \pi} \,
    \bar{\psi}_w(\vec x) \gamma_0
    \,\vec x \cdot \!\hat{z} \,
    S_F(\vec x, \vec y; E_1) \gamma_{\mu}
\nonumber \\ && \times \,
    S_F(\vec y, \vec z; E_2) \gamma^{\mu} \psi_v(\vec z)
    D_{\epsilon}(E_1 + \omega - \epsilon_w)
    D_{\epsilon}(E_2 + k_0 - \epsilon_v)
    D_{\epsilon}(E_2 + k_0 - E_1 ).
\end{eqnarray}
Replacing the first electron propagator with a spectral representation
gives
\begin{eqnarray}
S_{\rm SR} &=& -8 i\pi^3 \alpha \sum_m
      \int\! d^{\,3}x
    \!\int\! {dE_1 \over 2 \pi}
    \!\int\! {dE_2 \over 2 \pi} \,
    {\bar{\psi}_w(\vec x) \gamma_0 \,\vec x \cdot \!\hat{z}\, \psi_m(\vec x)
    \over E_1 - \epsilon_m(1-i\delta)} \, \Sigma_{mv}(E_2)
\nonumber \\ && \times \,
    D_{\epsilon}(E_1 + \omega - \epsilon_w)
    D_{\epsilon}(E_2 + k_0 - \epsilon_v)
    D_{\epsilon}(E_2 +k_0 - E_1).
\end{eqnarray}
The $D_{\epsilon}$ functions emphasize $E_1 = \epsilon_v$. If the state
$v$ is excluded in the sum over states $m$, the same kind of manipulations
applied to the vertex graph allow one to determine a ``perturbed orbital''
(PO) contribution from the SR diagram of
\begin{equation}
\delta z_{wv}({\rm PO};ns) = \!\int\! d^{\,3}x \!\sum_{m \neq v}
    {\bar{\psi}_w(\vec x) \gamma_0 \,\vec x \cdot \!\hat{z}\, \psi_m(\vec x)
    \over \epsilon_v - \epsilon_m + i \delta} \, \Sigma_{mv}(\epsilon_v).
\end{equation}
The notation $({\rm PO};ns)$ refers to the fact that this contribution is
essentially the self-energy $\Sigma_{\tilde{v}v}(\epsilon_v)$, where
$\tilde{v}$ is a perturbation of the $ns$ state. A similar contribution,
$\delta z_{wv}({\rm PO};np)$ arises from the side-left diagram, and is so
designated because it is $\Sigma_{w \tilde{w}}(\epsilon_w)$.
Its specific value is
\begin{equation}
\delta z_{wv}({\rm PO};np) = \!\int\! d^{\,3}x \!\sum_{m \neq w}\!
    \Sigma_{wm}(\epsilon_w)
    {\bar{\psi}_m(\vec x) \gamma_0 \,\vec x \cdot \!\hat{z}\, \psi_v(\vec x)
    \over \epsilon_w - \epsilon_m + i \delta}.
\end{equation}

The case when $m=v$ requires more care. We need to make a Taylor expansion
of the electron propagator in the self-energy function around the point
$E_2 = \epsilon_v - k_0$,
\begin{equation}
S_F(\vec y, \vec z; E_2) =
    S_F(\vec y, \vec z; \epsilon_v-k_0) + (E_2 - \epsilon_v + k_0)
    S'_F(\vec y, \vec z; \epsilon_v-k_0) + \,\cdots\,.
\label{TaylorS}
\end{equation}
The first term of the expansion is highly divergent, involving the integral
\begin{eqnarray}
&&   \int\! {dE_1 \over 2 \pi}
    \!\int\! {dE_2 \over 2 \pi} \,
    {1 \over E_1 - \epsilon_v + i \delta} \,
    D_{\epsilon}(E_1 + \omega - \epsilon_w)
    D_{\epsilon}(E_2 + k_0 - \epsilon_v)
    D_{\epsilon}(E_2 +k_0 - E_1)
\nonumber \\ &=&
    { 1 \over 8 i \pi^3 \,[ (\omega - \omega_0)^2 + \epsilon^2 ] } -
    \epsilon \left\{ { \omega - \omega_0 \over \pi^3 \,
    [ (\omega - \omega_0)^2 + 9 \epsilon^2 ] \,
    [ (\omega - \omega_0)^2 +   \epsilon^2 ] } \right\}.
\end{eqnarray}
Because we take the limit $\epsilon \to 0$ before $\omega \to \omega_0$
the second term can be dropped, leaving the divergent expression
\begin{equation}
\delta z_{wv}({\rm Div}) = - {i \over 2 \epsilon}
    z_{wv} \Sigma_{vv}(\epsilon_v)
\end{equation}
In the energy formalism this corresponds to a $1/\epsilon^2$ term that is
canceled by a term in the denominator of Eq.~(\ref{eq:eq1}). Here it does
not cancel, but instead forms the second term of the Taylor expansion of
the phase factor
\begin{equation}
e^{-i [\Sigma_{vv}(\epsilon_v) + \Sigma_{ww}(\epsilon_w)]/ 2 \epsilon}
\end{equation}
multiplying $z_{wv}$, where we have now included the effect of the
side-left diagram. Consideration of higher-order diagrams with two
self-energies and a photon interaction show that even more divergent
terms going as, for example, $\Sigma^2_{vv}(\epsilon_v)/\epsilon^2$
are present that continue the Taylor expansion. Thus this divergent term
is not directly canceled as in the energy formalism but instead can be
ignored because it enters only as a phase \cite{caveat}. However, the
second term in Eq.~(\ref{TaylorS}) does contribute a finite amount to
the scattering amplitude,
\begin{equation}
\delta z_{wv}({\rm D}) = {1 \over 2} \, z_{wv}
    \left[\Sigma'_{vv} (\epsilon_v) + \Sigma'_{ww}(\epsilon_w)\right],
\end{equation}
where we have included a similar term arising from the side-left diagram.
The treatment of these derivative (D) terms follows that of the vertex,
but in this case we simply sum the finite part of the free propagator
term, the Wick rotated term, and the terms in which poles are encircled
and present the results as $R_{wv}({\rm D})$ in the fourth row of
Table~\ref{tab:tab2}. The ultraviolet infinite part of the\
derivative terms cancels a similar term from the vertex exactly,
and the reference states are allowed to cancel numerically. Finally,
the two perturbed orbital terms are listed as $R_{wv}({\rm PO};ns)$
and $R_{wv}({\rm PO};np)$ in the fifth and sixth rows of
Table~\ref{tab:tab2}. The sums of all contributions give the
radiative correction results $R_{wv}$ shown in the last row of
Table~\ref{tab:tab2}.

\section{\label{sec:hions}Hydrogenic Ions}

In a previous paper \cite{SPC}, the imaginary part of the two-loop Lamb
shift was used to calculate radiative corrections to the $2p_{1/2}$ and
$2p_{3/2}$ lifetimes of hydrogenic ions. In this section, we redo the
$2p_{1/2}$ correction for one of these ions using the present matrix
element formalism as a check of both approaches and to show the role of
radiative energy shifts in lifetime corrections. The $2p_{3/2}$ case
included M1 decays, and so cannot be used for a direct comparison.

In Ref.~\cite{SPC}, the radiative correction to the decay rate $\Gamma$
is defined in terms of the function $R(Z\alpha)$ by
\begin{equation}
\Gamma = \Gamma_0 \left[ 1 + {\alpha \over \pi}\, R(Z\alpha) \right].
\end {equation}
An important feature of these corrections is that both radiative
corrections to energies and radiative corrections to matrix elements
are of similar importance. As E1 decay rates in the length gauge are
the product of the cube of the energy difference $\omega_0$ and the
square of the transition matrix element $z_{wv}$, one component of
$R(Z\alpha)$ which arises from energy corrections should be given by
\begin{equation}
3 \, {\delta E \over E} = {\alpha \over \pi} \, (Z \alpha)^2 \!
    \left[ -{32 \over 3} \, {\rm ln}(Z\alpha)^{-2} + 22.815 \right]
\end{equation}
where we have used the standard values of the leading $Z\alpha$-expansion
contributions to the $1s$ and $2p_{1/2}$ Lamb shifts. At $Z = 5$, this
contributes $-0.064$ to $R(Z\alpha)$. However, the actual value of this
function calculated in Ref.~\cite{SPC} was $-0.014$, so that a substantial
positive contribution of about 0.050 must come from the shift in the
matrix element. We have carried out this calculation using the techniques
described above and find a result of 0.025 for the function $R_{wv}$
defined in Eq.~(\ref{eq:rwv}). This result is in perfect agreement with
the expected value, as it has to be doubled to account for
the fact that the matrix element enters as a square in decay rate
calculations. But while this result shows the consistency between the
present matrix-element and the former decay-rate approaches, there is
no improvement in the numerical accuracy using our present method
for these radiative correction calculations.

\section{\label{sec:results}Results and Discussion}

As found with our previous work on E1 matrix elements for hydrogenic
ions \cite{SPC}, there is a high degree of cancellation between
contributing terms shown in the first six rows of Table~\ref{tab:tab2}.
This is due to the fact that radiative corrections to decay rates
enter in order $\alpha (Z \alpha)^2$. In fact, if one puts the
perturbed orbital contributions aside, the cancellation is almost
complete. As a result, the radiative correction is dominated by the
$({\rm PO},ns)$ terms, which are the larger of the two perturbed
orbital terms. Since it is a standard practice with the alkalis to
derive matrix elements from observed lifetimes by dividing out the
cube of the experimental energy differences, we do not need to
consider radiative corrections to energies here, as they are
automatically excluded in empirically extracted matrix elements.

The experimental accuracies for the alkali matrix elements are half the
accuracies of the lifetimes quoted in the introduction, specifically being
130 ppm, 490ppm, 940ppm, 1440ppm, 50 ppm, and 1870 ppm for lithium through
francium. Comparing with the results of Table~\ref{tab:tab2}, we see
corresponding theoretical contributions of 12 ppm, 39ppm, 149ppm, 488ppm,
606ppm, and 2158 ppm. Thus these contributions are in principle measurable
for several of the alkalis, particularly cesium and francium.

For the case of lithium, the present approach is certainly less accurate
than that available from nonrelativistic quantum electrodynamics (NRQED).
The nonrelativistic lithium wavefunction is extremely well understood
using variational techniques \cite{liv}, and the same kind of NRQED
calculations reported in Ref.~\cite{SPC} for hydrogenic ions can
certainly be used for neutral lithium. A nonrelativistic evaluation of
the lifetime of the lithium $2p_{1/2}$ state has been carried out in
Ref.~\cite{Yan} with a result of 27.1045(14) ns which includes finite
mass corrections along with relativistic corrections. The quoted error
came mainly from the fact that the relativistic corrections were not
directly calculated, and could be eliminated with rigorous relativistic
calculations. Regardless, the radiative correction calculated here is
so small that it should not make any difference to the theoretical
lifetime in comparing with experiment.

The cases of cesium and francium are more problematic because of the
complexity of their wave functions. A great deal of effort has gone into
treating these wave function accurately, largely spurred by interest in
PNC transitions. However, while experimental accuracy is now certainly
high enough for detection of radiative corrections of the size found here
for cesium and almost at the level needed for francium, further advances
in many-body theory will be required, as was the case for the Lamb shift
\cite{neutrallamb} and the hyperfine splitting \cite{HFS}, before one
can decisively say this effect has been seen. Incidentally, a theoretical
advantage for lifetimes as compared to hyperfine splittings is the
relative insensitivity of lifetimes to nuclear effects: as emphasized in
Ref.~\cite{HFS}, uncertainties in the distribution of nuclear magnetism
lead to theoretical uncertainties that are difficult to control.

\begin{figure}[p]
\centerline{\includegraphics[scale=0.6]{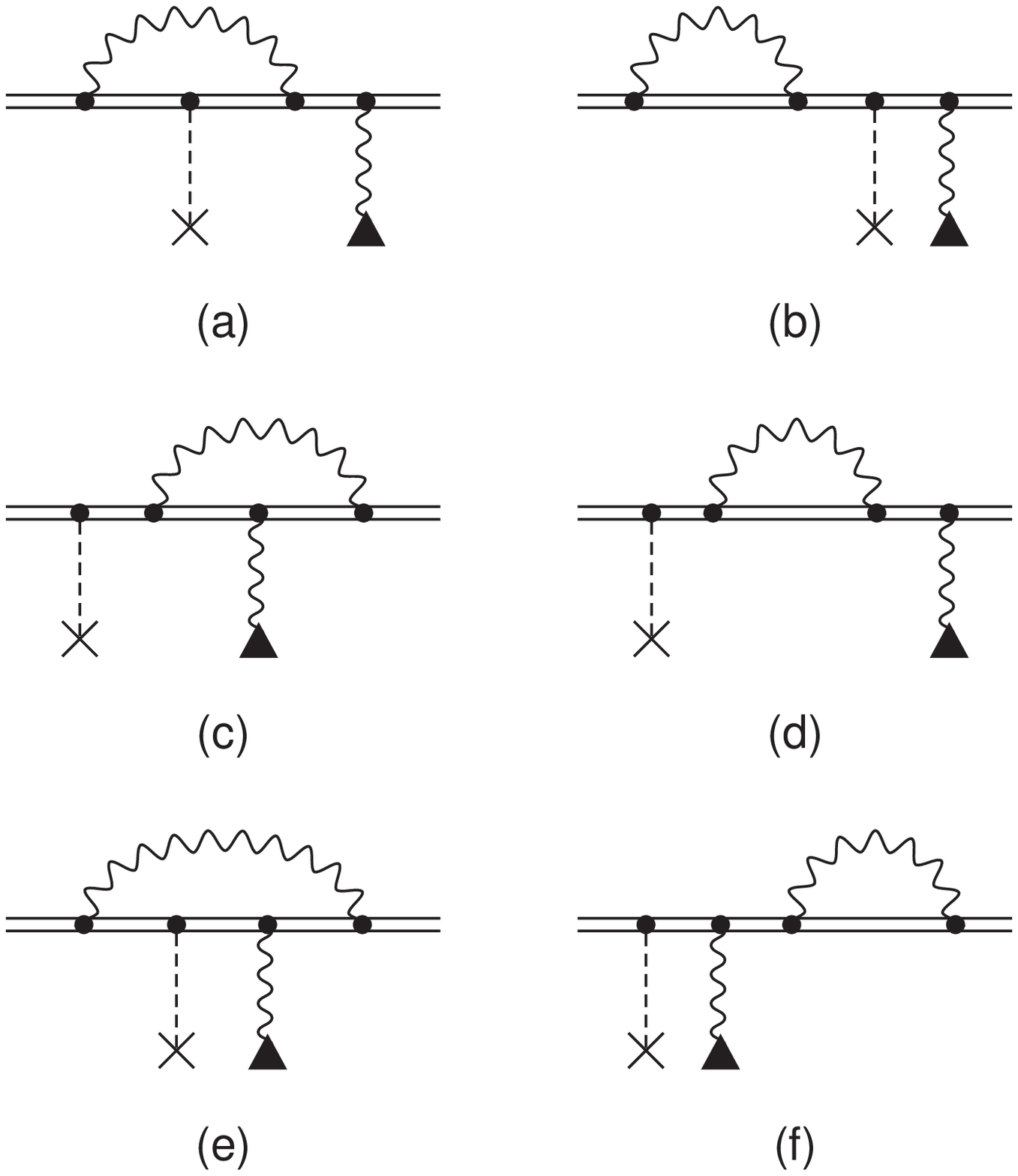}}
\caption{\label{fig:fig1}
Feynman diagrams for the radiative correction to electron excitation by
a laser photon, indicated by the wavy line terminated with a triangle,
in the presence of interaction with the nucleus through the exchange of
a Z boson, indicated by the dashed line terminated with a cross.}
\end{figure}

\begin{figure}[p]
\centerline{\includegraphics[scale=0.6]{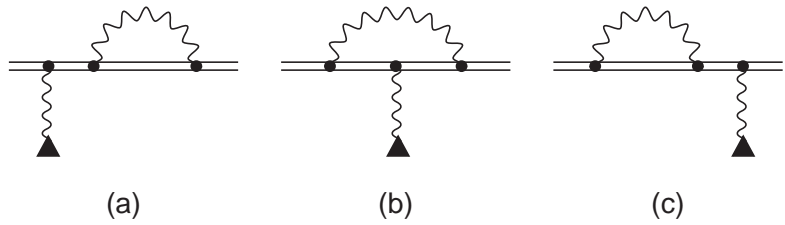}}
\caption{\label{fig:fig2}
Feynman diagrams for the radiative correction to the matrix element for
ns + photon $\rightarrow$ np.}
\end{figure}

A major spur to theoretical work on accurate cesium wave functions is the
fact that the observation of PNC in the atom \cite{Wieman} has significant
implications for particle physics. Before the large binding correction
to the lowest-order radiative correction $-\alpha/2\pi$ was found,
a discrepancy with experiment existed. As discussed in the introduction,
a motivation for the present work was to provide a basis for the full
calculation of the diagrams shown in Fig.~\ref{fig:fig1}. We note,
for example, that diagram~\ref{fig:fig1}c is of the same form as the
vertex diagram V in Fig.~\ref{fig:fig2}b, with the difference that the
$6s$ (or $6p_{1/2}$) state in diagram~\ref{fig:fig2}b must be replaced
with a perturbed orbital of the same parity which arises from the
$6p_{1/2}$ (or $6s$) state perturbed by a Z boson exchange with the
nucleus. Thus a relatively straightforward modification of the codes
developed for the present calculations will allow the determination of
this term. In fact, with the exception of diagram~1e, the entire PNC
calculations of Fig.~\ref{fig:fig1} can be treated as perturbations
of either the weak interaction calculations \cite{pncrad3}, the
self-energy calculations \cite{neutrallamb}, or the present transition
matrix calculations. We are thus in a position to carry out the bulk
of the full radiative correction calculations. The fact that the
previously neglected radiative correction to the E1 matrix element
is $0.261 \alpha / \pi$ as shown in Table~\ref{tab:tab2} makes it
likely that the full PNC calculation will differ quantitatively,
though perhaps not qualitatively, from the calculations that
consider only the Z vertex.

Since the lowest-order results presented in Table~\ref{tab:tab1} are
substantially corrected by higher-order effects in many-body perturbation
theory (MBPT) \cite{Idrees}, similar corrections can be expected for the
radiative corrections. The simplest way to account for these corrections
is to assume that $R_{wv}$, defined as a ratio of the radiative
correction to the lowest-order matrix element, is valid when the
lowest-order dipole matrix element is replaced with more accurate results
obtained with MBPT methods. As the size of the radiative corrections
is small, even if this is only roughly true, the basic size of the
effect would have been established here. To attain more accuracy,
a QED perturbation theory approach could be taken. In that case,
one would first consider graphs of the type shown in Fig.~\ref{fig:fig1}
where the Z vertex is replaced by interactions with other electrons.
For even higher accuracy, one could consider yet more complicated
graphs with one absorbed photon, one radiative photon, and two
interactions with the other electrons: this is known, in the case
without the radiative photon, to give results within a few percent of
the experimental answers \cite{Idrees}. Progress in these large scale
calculations, taken together with continuing advances in experiment
and many-body methods, should allow tests of the lifetimes of the
alkalis at the level achieved for positronium.

\acknowledgments
The work of J.S.\ was supported in part by NSF grant PHY-0097641.
The work of K.T.C.\ was performed under the auspices of the
U.S.\ Department of Energy at the University of California,
Lawrence Livermore National Laboratory under Contract No. W-7405-ENG-48.
We thank Steve Libby for discussions.

\newpage

%%% ***** Table 1 *****
%%%
\begin{table*}[t]
\caption{\label{tab:tab1}
Dipole matrix elements $r_{wv}$ for $ns_{1/2}-np_{1/2}$ and
$ns_{1/2}-np_{3/2}$ transitions in the alkalis with Kohn-Sham
potentials. Units a.u..}
\begin{ruledtabular}
\begin{tabular}{ccccccc}
transition          &   Li  &   Na  &   K   &   Rb  &   Cs  &   Fr  \\
\colrule
$ns_{1/2}-np_{1/2}$ & 4.171 & 4.588 & 5.681 & 6.009 & 6.585 & 6.511 \\
$ns_{1/2}-np_{3/2}$ & 4.171 & 4.587 & 5.679 & 5.995 & 6.545 & 6.328 \\
\end{tabular}
\end{ruledtabular}
\end{table*}

%%% ***** Table 2 *****
%%%
\begin{table*}[b]
\caption{\label{tab:tab2}
Self-energy contributions $R_{wv}$ to E1 matrix elements for
$ns_{1/2} - np_{1/2}$ transitions in the alkalis: error of 0.001 for Sum.
Units $(\alpha / \pi) z_{wv}$.}
\begin{ruledtabular}
\begin{tabular}{lrrrrrr}
   \multicolumn{1}{c}{Term}
& \multicolumn{1}{c}{Li}
& \multicolumn{1}{c}{Na}
& \multicolumn{1}{c}{K}
& \multicolumn{1}{c}{Rb}
& \multicolumn{1}{c}{Cs}
& \multicolumn{1}{c}{Fr} \\
\colrule
$R_{wv}({\rm V};00)$   &  -9.162 &  -9.308 &  -9.509 &  -9.573 & 
-9.655 &  -9.647 \\
$R_{wv}({\rm V};iw)$   & 134.712 & 105.062 & 127.516 & 125.136 & 
132.719 & 112.464 \\
$R_{wv}({\rm V;Poles})$&-137.256 &-107.481 &-129.741 &-127.273 
&-134.877 &-114.624 \\
$R_{wv}({\rm D})$      &  11.698 &  11.712 &  11.731 &  11.738 & 
11.748 &  11.747 \\
$R_{wv}({\rm PO};ns)$  &   0.003 &   0.031 &   0.067 &   0.182 & 
0.326 &   0.787 \\
$R_{wv}({\rm PO};np)$  &   0.000 &   0.001 &   0.000 &   0.000 & 
0.000 &   0.202 \\
[4pt] \colrule
Sum                    &  -0.005 &   0.017 &   0.064 &   0.210 & 
0.261 &   0.929 \\
\end{tabular}
\end{ruledtabular}
\end{table*}


\begin{thebibliography}{99}
%
  \newpage
%
\bibitem{Vallery} R.S. Vallery, P.W. Zitzewitz, and D.W. Gidley,
    Phys. Rev. Lett. {\bf 90}, 203402 (2003).
\bibitem{Gidley} A.H. Al-Ramadhan and D.W. Gidley,
    Phys. Rev. Lett. {\bf 72}, 1632 (1994).
\bibitem{liexp} W.I. McAlexander, E.R.I. Abraham, and R.G. Hulet,
    Phys. Rev. A {\bf 54}, R5 (1996).
\bibitem{naexp} K.M. Jones, P.S. Julienne, P.D. Lett, W.E. Phillips,
    E. Tiesinga, and C.J. Williams, Europhys. Lett. {\bf 35}, 85 (1996).
\bibitem{kexp}
    H. Wang, J. Li, T. Wang, C.J. Williams, P.L. Gould, and W.C. Stwalley,
    Phys. Rev. A {\bf 55}, R1569 (1997).
\bibitem{rbexp}
    R.F. Gutterres, C. Amiot, A. Fioretti, C. Gabbanini, M. Mazzoni, and
    I. Dulieu, Phys. Rev. A {\bf 66}, 024502 (2002).
\bibitem{csexp}
    C. Amiot, O. Dulieu, R.F. Gutterres, and F. Masnou-Seeuws,
    Phys. Rev. A {\bf 66}, 052506 (2002).
\bibitem{frexp} J.E. Simsarian, L.A. Orozco, G.D. Sprouse, and W.Z. Zhao,
    Phys. Rev. A {\bf 57}, 2448 (1998).
\bibitem{SPC} J. Sapirstein, K. Pachucki and K.T. Cheng,
    Phys. Rev. A {\bf 69}, 02113 (2004).
\bibitem{Wieman}
    S.C. Bennett and C.E. Wieman, Phys. Rev. Lett. {\bf 82}, 2484 (1999);
    C.S. Wood {\it et. al.}, Science {\bf 275}, 1759 (1997).
\bibitem{pncrad1} M. Yu. Kuchiev, J. Phys. B {\bf 35}, L503 (2002).
\bibitem{pncrad2} A.I Milstein, O.P. Sushkov, and I.S. Terekhov,
    Phys. Rev. Lett. {\bf 89}, 283003 (2002).
\bibitem{pncrad3} J. Sapirstein, K. Pachucki, A. Veitia, and K.T. Cheng,
    Phys. Rev. A {\bf 67}, 052110 (2003).
%\bibitem{Gould} Jason M. Amini and Harvey Gould,
%   Phys. Rev. Lett. {\bf 91}, 153001 (2003).
%\bibitem{Derevianko} A. Derevianko and S.G. Porsev,
%   Phys. Rev. A {\bf 65}, 053403 (2002).
\bibitem{neutrallamb} J. Sapirstein and K.T. Cheng,
    Phys. Rev. A {\bf 66}, 042501 (2002).
\bibitem{HFS} J. Sapirstein and K.T. Cheng,
    Phys. Rev. A {\bf 67}, 022512 (2003).
\bibitem{liv}
    K. Pachucki and J. Komasa, Phys. Rev. A {\bf 68}, 042507 (2003);
    Zong-Choa Yan and G.W.F. Drake, Phys. Rev. A {\bf 52}, R4316 (1995);
    F. King, Phys. Rev. A {\bf 40}, 1735 (1989).
\bibitem{Sucher} J. Sucher, Phys. Rev. {\bf 107}, 1448 (1957).
\bibitem{BCS} S.A. Blundell, K.T. Cheng, and J. Sapirstein,
    Phys. Rev. A {\bf 55}, 1857 (1997).
%\bibitem{JS} W.R. Johnson and G. Soff,
%   At. Data and Nucl. Data Tables {\bf 33}, 405 (1985).
%\bibitem{Derevianko} A. Derevianko, Phys. Rev. A {\bf 65}, 012106 (2000).
\bibitem{caveat} The fact that the self energy also has an imaginary part
    associated with decay is under further analysis; while this term would
    not cancel as a phase, in this formalism we note that the states are
    stable at large positive and negative times, and this fact may
    justify neglecting the imaginary part. See also the discussion in
    section 4 of  U.D. Jentschura, C.H. Keitel,  and K. Pachucki, Can. J. 
    Phys. {\bf 80}, 1213 (2002).
\bibitem{Yan} Zong-Chao Yan, M. Tambasco, and G.W.F. Drake,
    Phys. Rev. A {\bf 57}, 1652 (1998).
\bibitem{Idrees} W.R. Johnson, M. Idrees, and J. Sapirstein,
    Phys. Rev. A {\bf 35}, 3218 (1987).
\end{thebibliography}
\end{document}